\documentclass[multphys,vecphys]{svmult}

\usepackage{makeidx}         
\usepackage{graphicx}        
\usepackage{multicol}        
\usepackage[bottom]{footmisc}

\makeindex             

\begin{document}

\title*{The role of AGN feedback and gas viscosity in hydrodynamical
  simulations of galaxy clusters}
\titlerunning{Simulations of AGN feedback and gas viscosity in
  clusters} 
\author{Debora Sijacki \and 
Volker Springel}
\institute{Max-Planck-Institut f\"{u}r Astrophysik,
  Karl-Schwarzschild-Stra\ss{}e 1, \\85740 Garching bei M\"{u}nchen,
  Germany \texttt{deboras@mpa-garching.mpg.de}
}
%
%
\maketitle

\begin{abstract}
We study the imprints of AGN feedback and physical viscosity on the
properties of galaxy clusters using hydrodynamical simulation models
carried out with the TreeSPH code {\small GADGET-2}. Besides
self-gravity of dark matter and baryons, our approach includes
radiative cooling and heating processes of the gas component and a
multiphase model for star formation and SNe feedback
\cite{sijacki:SH03}. Additionally, we introduce a prescription for
physical viscosity in {\small GADGET-2}, based on a SPH discretization
of the  Navier-Stokes and general heat transfer equations. Adopting
the Braginskii parameterization for the shear viscosity coefficient,
we explore how gas viscosity influences the properties of AGN-driven
bubbles. We find that the morphology and dynamics of bubbles are
significantly affected by the assumed level of physical viscosity in
our simulations, with higher viscosity leading to longer survival
times of bubbles against fluid instabilities. In our cosmological
simulations of galaxy clusters, we find that the dynamics of mergers
and the motion of substructures trough the cluster atmosphere is
significantly affected by viscosity. We also introduce a novel,
self-consistent AGN feedback model where we simultaneously follow the
growth and energy release of massive black holes embedded in a cluster
environment. We assume that black holes accreting at low rates with
respect to the Eddington limit are in a radiatively inefficient
regime, and that most of the feedback energy will appear in a
mechanical form. Thus, we introduce AGN-driven bubbles into the ICM
with properties, such as radius and energy content, that are directly
linked to the black hole physics. This model leads to a self-regulated
mechanism for the black hole growth and overcomes the cooling flow
problem in host halos, ranging from the scale of groups to that of
massive clusters.
\end{abstract}

\section{Physical viscosity in SPH simulations of galaxy clusters}
\label{sijacki:Viscosity}
There is growing observational evidence \cite{sijacki:Markevitch2002,
  sijacki:Sun2006, sijacki:Fabian2006} that gas viscosity in massive,
  hot clusters might not be negligible, and that it could play an
  important role in dissipating energy generated by AGN-driven bubbles
  or during merger events.
\subsection{Numerical implementation}
\label{sijacki:Viscosity_Implementation}
Within the framework of the entropy conserving formulation of SPH
\cite{sijacki:SH2002} in {\small GADGET-2} \cite{sijacki:Springel2005,
sijacki:Springel2001}, we have implemented a treatment of physical
viscosity that accounts both for the shear and bulk part, as explained in
detail in \cite{sijacki:SijackiSpringel2006b}. In particular, we have
derived novel SPH formulations of the Navier-Stokes and general heat
transfer equations, and for the shear viscosity coefficient we have
adopted Braginskii's parameterization \cite{sijacki:Braginskii1958,
sijacki:Braginskii1965}. We have tested our numerical scheme
extensively on a number of hydrodynamical problems with known
analytic solutions, recovering these solutions accurately.

\subsection{AGN-driven bubbles in a viscous ICM}
\label{sijacki:Viscosity_bubbles}
\begin{figure}
\centering
\vspace{-0.6truecm}
\includegraphics[height=4.5cm]{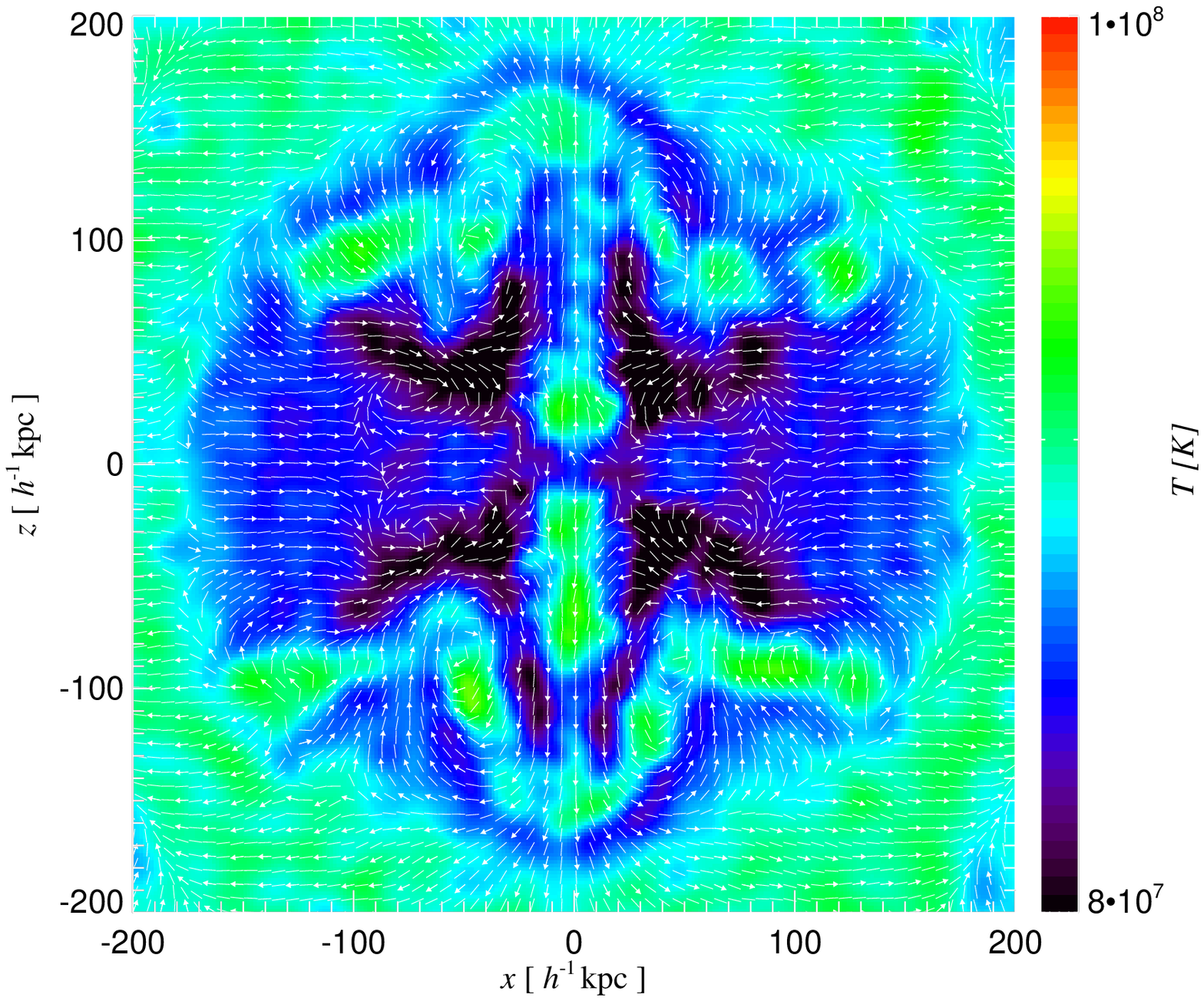}
\hspace{0.3truecm}
\includegraphics[height=4.5cm]{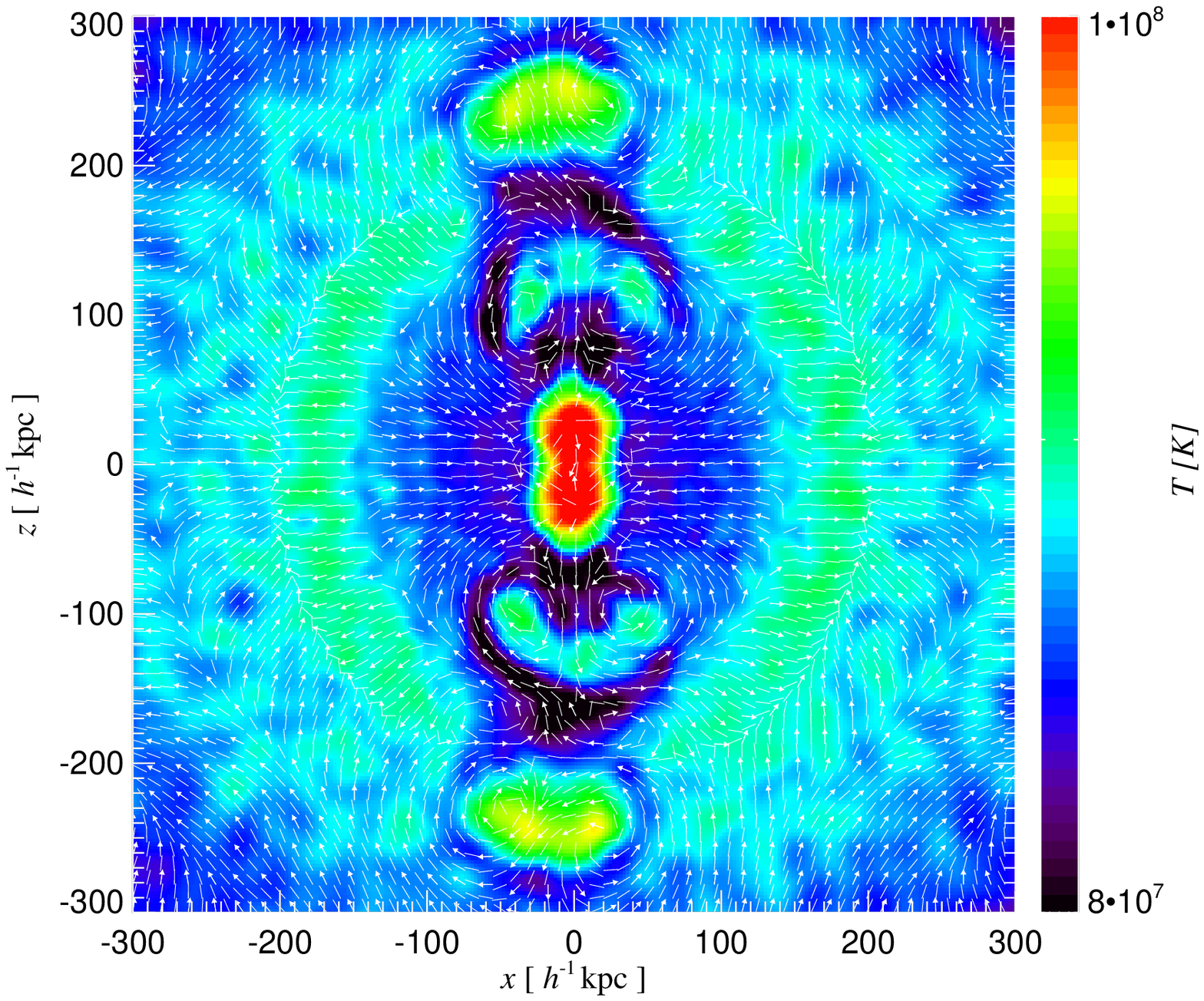}
\vspace{-0.4truecm}
\caption{Mass-weighted temperature maps of a $10^{15}\,h^{-1} {\rm M_\odot}$
  isolated halo, subject to AGN bubble heating. The velocity field of the
  gas is over-plotted with white arrows.}
\label{sijackiF1}  
\end{figure}

As a first application of our physical viscosity implementation in
{\small GADGET-2}, we have analyzed AGN-induced bubbles in a viscous
intracluster gas. We consider models of isolated galaxy clusters
consisting of a static NFW dark mater halo with a gas component which
is initially in hydrostatic equilibrium. AGN heating has been
simulated following a phenomenological approach, outlined in
\cite{sijacki:SijackiSpringel2006a}, where the bubbles are recurrently
injected in the central cluster region. In Fig.~\ref{sijackiF1} we
show temperature maps of a $10^{15}\,h^{-1} {\rm M}_\odot$ galaxy
cluster that is subject to AGN heating and has a certain level of
physical viscosity. In the left-hand panel, the Braginskii viscosity
has been suppressed by a factor of 0.3, while in the right-hand panel,
the simulation has been evolved with the full Braginskii viscosity. It
can be seen that the morphologies, maximum clustercentric distance
reached and survival times depend strongly on the assumed level of
physical viscosity. With unsuppressed Braginskii viscosity, bubbles
rise up to $\sim 300 h^{-1}{\rm kpc}$ in the cluster atmosphere
without being disrupted, and up to $2-3$ bubble episodes can been
detected, indicating that the bubbles survive as long as $\sim
2\times10^8$yr. However, in the case of suppressed physical viscosity
by a factor of $0.3$ bubbles start to disintegrate at roughly $150
h^{-1}{\rm kpc}$.

\subsection{Cosmological simulations of viscous galaxy clusters}
\label{sijacki:Viscosity_clusters}
\begin{figure}
\centering
\vspace{-0.6truecm}
\includegraphics[height=4.5cm]{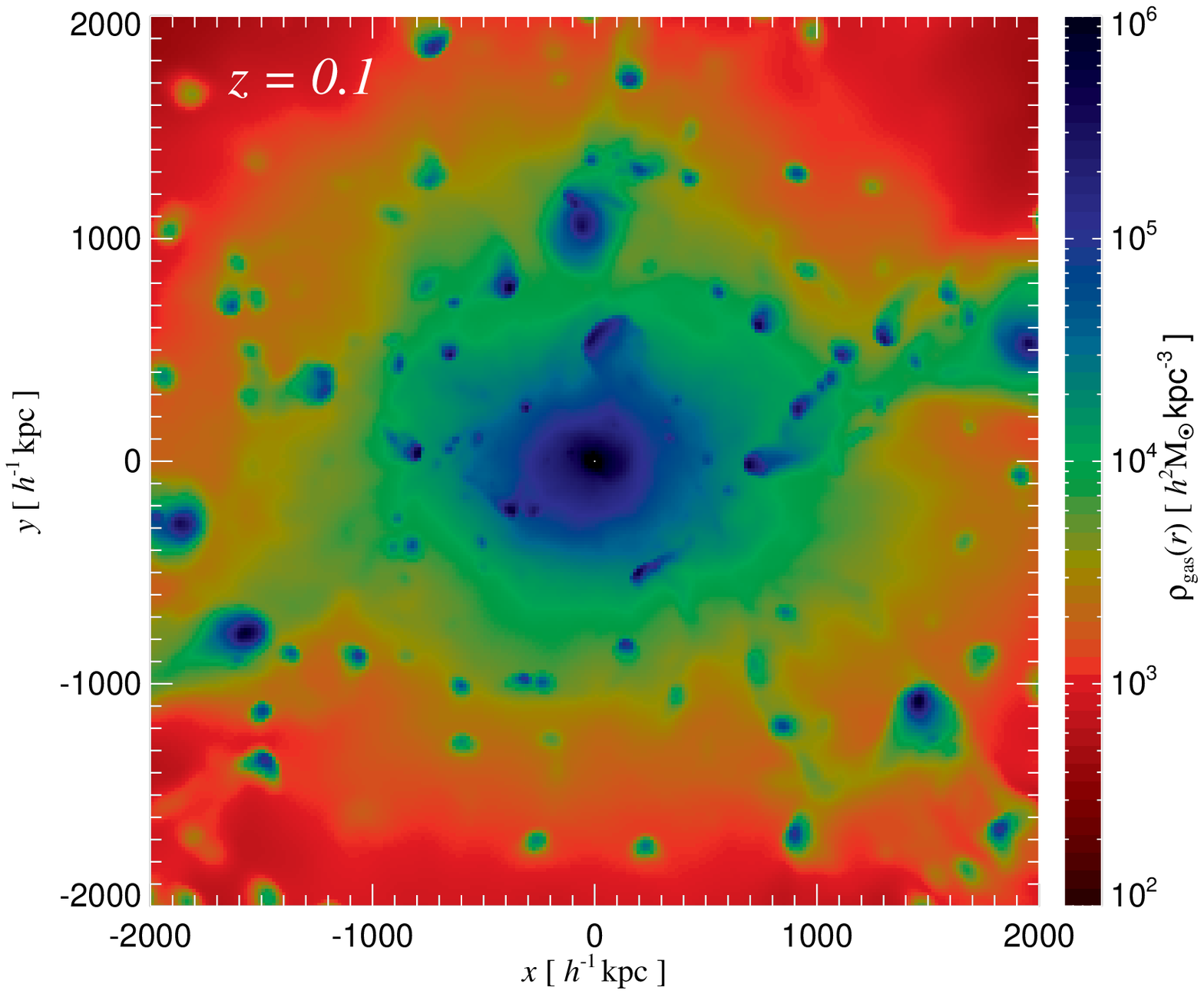}
\hspace{0.3truecm}
\includegraphics[height=4.5cm]{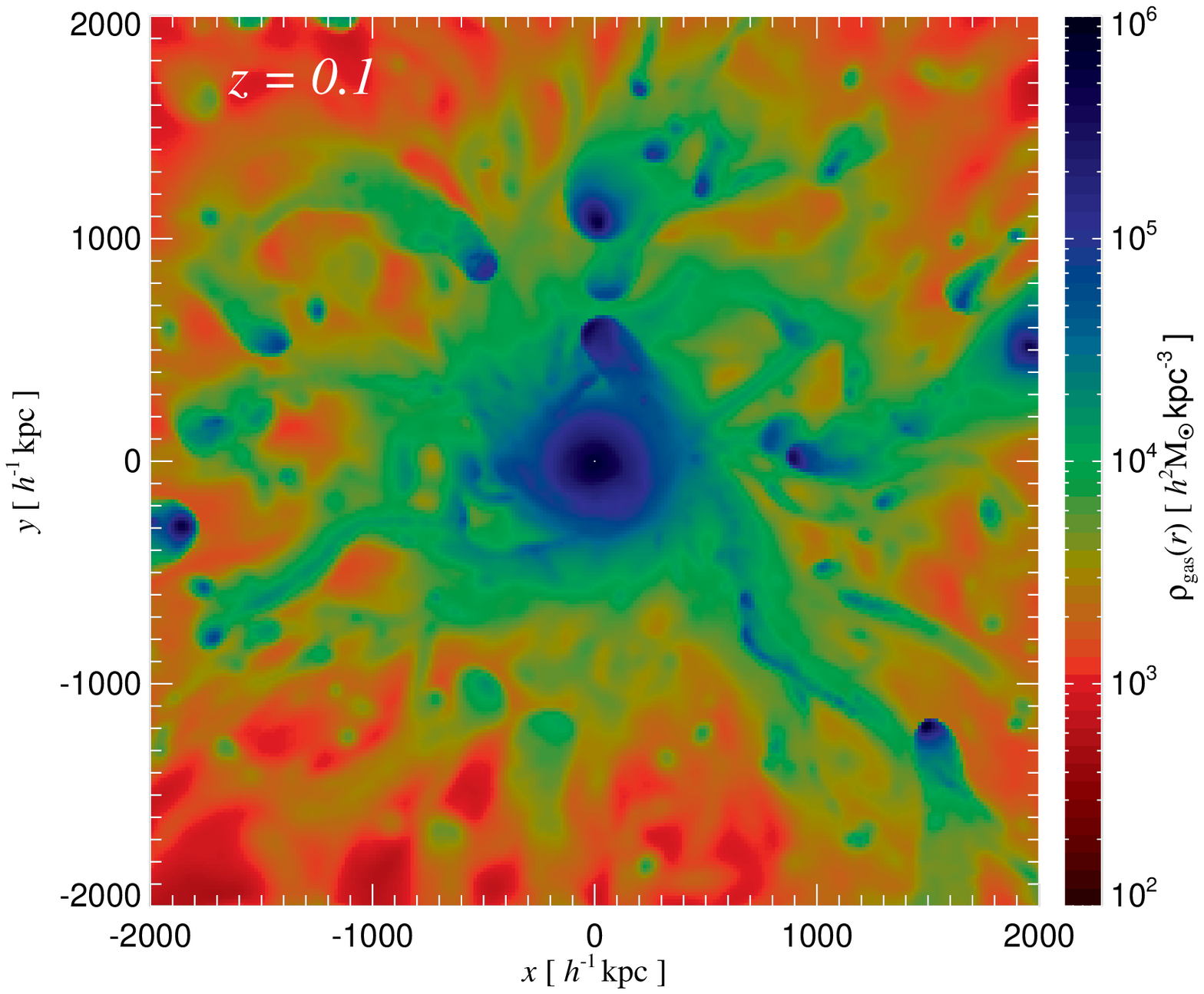}
\vspace{-0.4truecm}
\caption{Projected gas density maps of a galaxy cluster
  simulation at redshift $z=0.1$, as indicated in the
  upper-left corner of the panels. The left-hand panel shows the gas
  density distribution in the case of a non-radiative run, while the
  right-hand panel gives the gas density distribution when Braginskii shear
  viscosity is ``switched-on'', using a suppression factor of 0.3.}
\label{sijackiF2}  
\end{figure}

We have carried out fully self-consistent cosmological simulations of
galaxy clusters with certain amounts of physical viscosity. We have
both performed viscous non-radiative simulations and runs with
additional cooling and star formation, in order to understand the
complex interplay of these different physical ingredients. In
Fig.~\ref{sijackiF2} we show density maps of a non-radiative galaxy
cluster simulation at $z=0.1$ without any physical viscosity
(left-hand panel) and with 0.3 of Braginskii shear viscosity
(right-hand panel). We find that the introduction of a modest level of
physical viscosity has a significant impact on galaxy cluster
properties. The dynamics of clusters during merging events is
affected, with viscous dissipation processes generating an entropy
excess in cluster peripheries. Also, due to the viscous dissipation,
smaller structures entering more massive halos are more efficiently
stripped of their gaseous content, which forms narrow and up to $100
h^{-1}{\rm kpc}$ long tails, as visible on the right-hand panel of
Fig.~\ref{sijackiF2}. These features of viscous dissipation are very
prominent, occurring already at quite early cosmic times and
regardless of the presence of radiative cooling in the runs. However,
even though viscous dissipation occurs in the central cluster region,
it does not provide sufficient heating to prevent the formation of a
central cooling flow  at low redshifts, indicating that at least one
other physical process is needed to reconcile observational findings
with simulations.

\section{Self-regulated AGN feedback in simulations of galaxy clusters}
\label{sijacki:AGN}

\subsection{Methodology}
\label{sijacki:AGN_Methododlogy}
We have developed a novel AGN feedback model that simultaneously
tracks the growth of massive black holes in cluster environments, and
provides heating in the form of AGN-driven bubbles
\cite{sijacki:SijackiSpringel2006c}. In our simulations of clusters a
self-regulated feedback loop is established, leading to an equilibrium
where black hole growth is restricted and the central ICM is heated,
overcoming the cooling flow problem.

We model the black hole growth according to the prescriptions outlined
in \cite{sijacki:DiMatteo2005, sijacki:Springel2005b}. In these
studies, the Bondi formula has been adopted for the accretion rate
onto a black hole, and the Eddington limit has been imposed. Here, we
link the black hole properties, namely its  mass and accretion rate,
with the physics of AGN-blown bubbles. We parameterize our scheme in
terms of bubble energy and radius, and we consider recurrent episodes
of bubble injection. Specifically, we introduce a threshold in black
hole accreted mass above which a bubble event is triggered. We relate
the bubble energy with the black hole properties as follows,
\begin{equation}
E_{\rm bub} = f\, \epsilon_r \, c^2 \,\delta M_{\rm BH}\,,
\end{equation}  
where $f$ is the fraction of energy that goes into the
bubbles\footnote{For low accretion rates, $\dot M_{\rm BH} <
10^{-2}\dot M_{\rm Edd}$, we assume that most of the energy is in
mechanical form.}, $\epsilon_r$ is the standard radiative efficiency
that we assume to be $0.1$, and $\delta M_{\rm BH}$ is the mass growth
of a black hole between two successive bubble episodes. Moreover, we
link the bubble radius both to $\delta M_{\rm BH}$ and to the density
of the surrounding ICM, in the following way
\begin{equation}
R_{\rm bub} = R_{{\rm bub},0} \bigg(\frac{E_{\rm bub}/E_{{\rm
bub},0}}{\rho_{\rm ICM}/\rho_{{\rm ICM},0}} \bigg)^{1/5}\,,
\end{equation}  
where $R_{{\rm bub},0}$, $E_{{\rm bub},0}$, and $\rho_{{\rm ICM},0}$
  are normalization values for the bubble radius, energy content and
  ambient density. The relation for the bubble radius is motivated by
  the solutions for the radio cocoon expansion in a spherically
  symmetric case \cite{sijacki:Begelman1989}.

\subsection{Black hole growth and feedback in isolated galaxy clusters}
\label{sijacki:AGN_Isolated}
\begin{figure}
\centering 
\vspace{-0.4truecm}
\includegraphics[height=5.8cm]{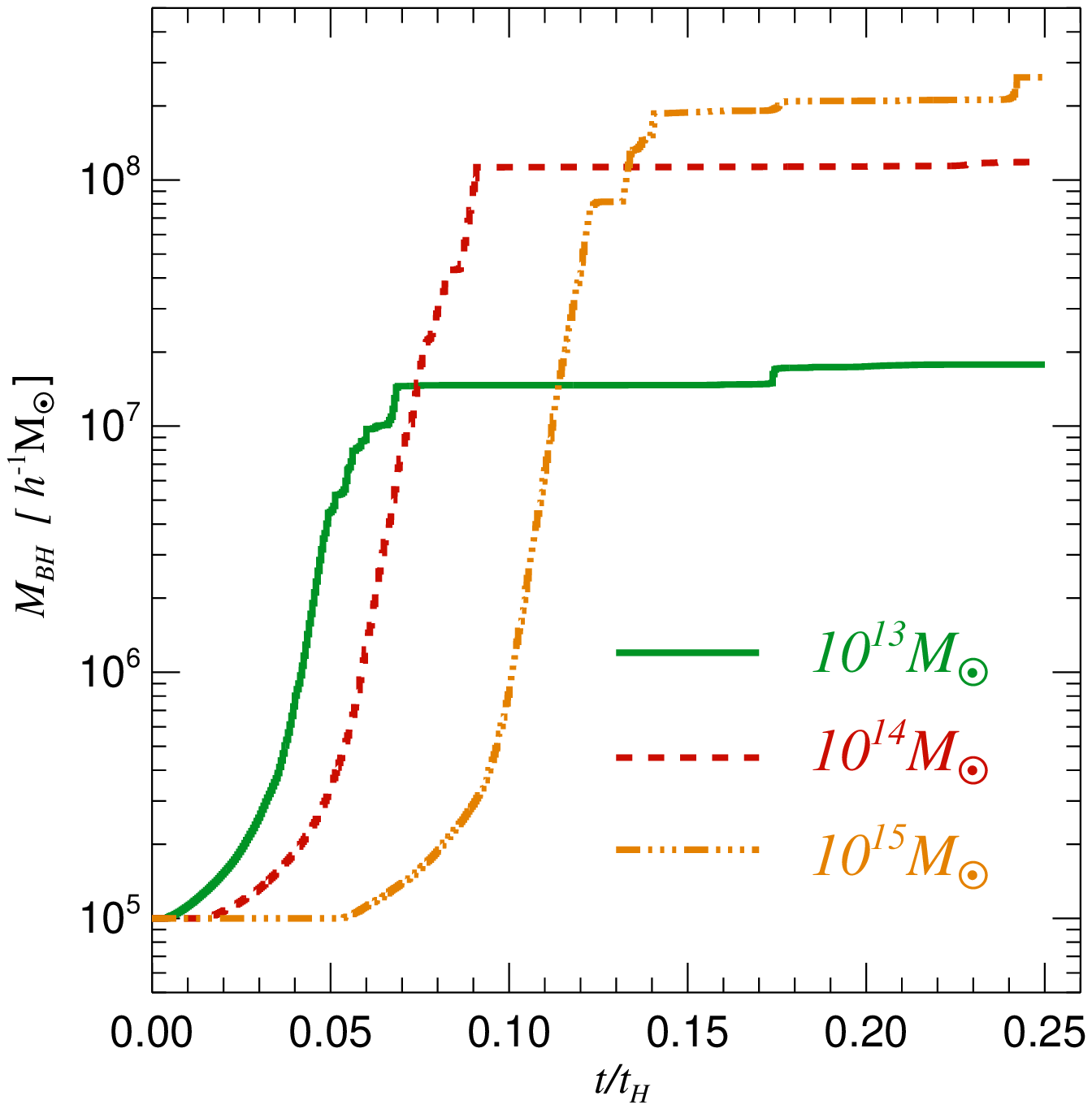}
\hspace{-0.3truecm} 
\includegraphics[height=5.8cm]{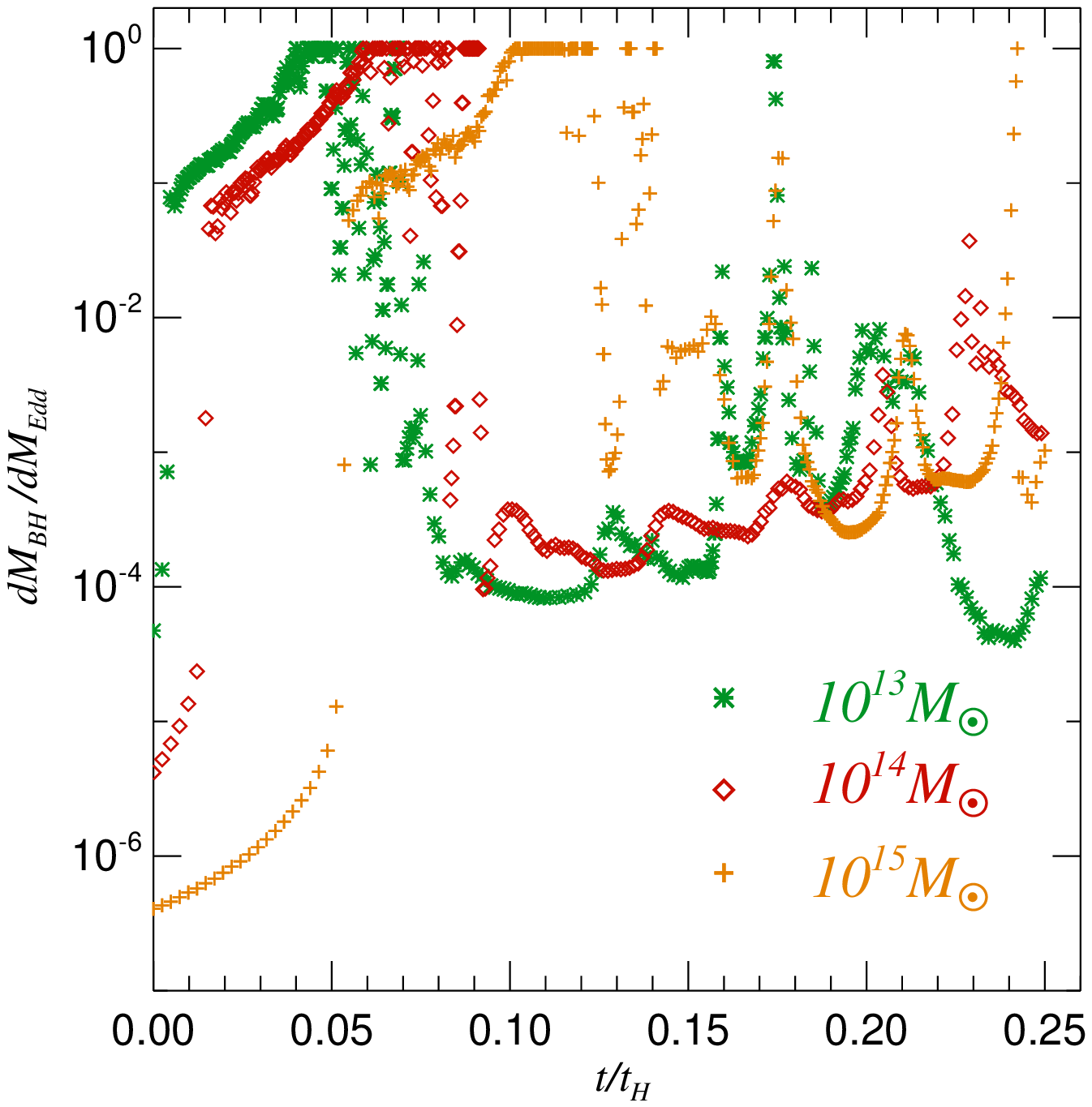}
\vspace{-0.4truecm}
\caption{Black hole mass growth and black hole accretion rate in
  Eddington units for three isolated galaxy clusters
  simulations. After the initial rapid growth from a $10^{5}\,h^{-1}
  {\rm M}_\odot$ seed, the black hole mass is regulated and the
  accretion rate drops to low values.}
\label{sijackiF3}  
\end{figure}

\begin{figure}
\centering 
\vspace{-0.4truecm}
\includegraphics[height=5.8cm]{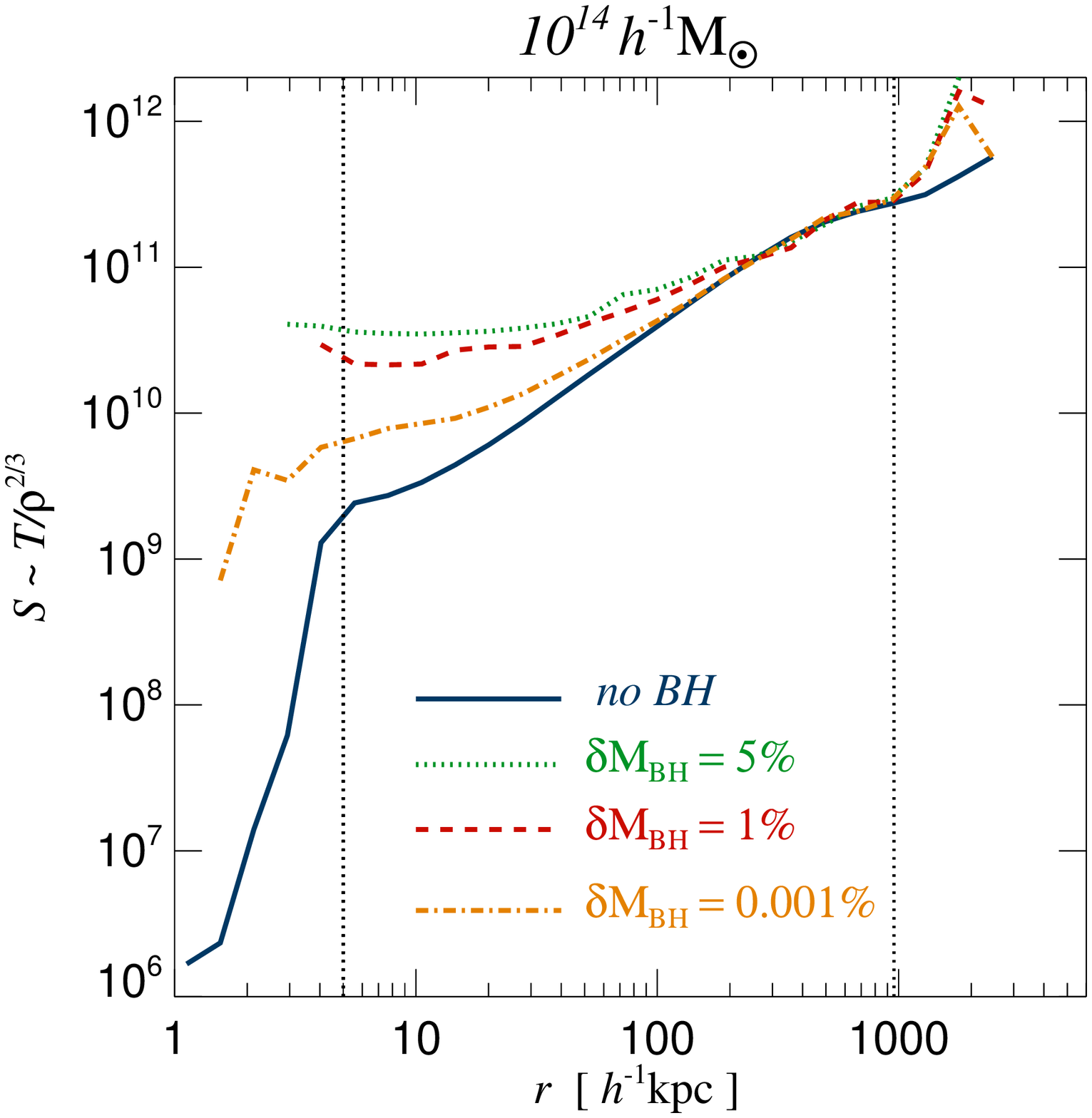}
\hspace{-0.3truecm} 
\includegraphics[height=5.8cm]{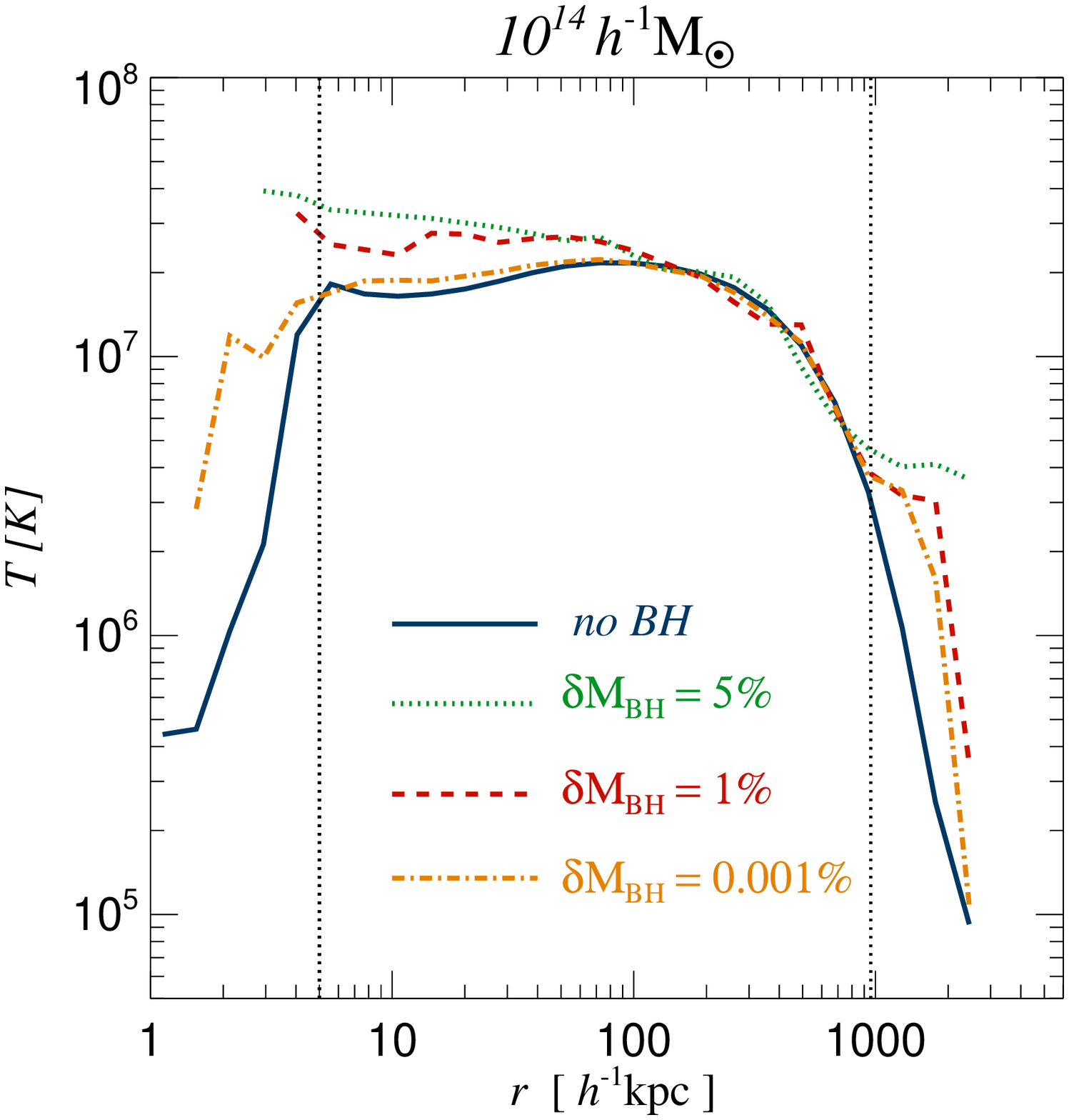}
\vspace{-0.4truecm}
\caption{Entropy (left-hand panel) and mass-weighted temperature
  (right-hand panel) radial profiles of a $10^{14}\,h^{-1} {\rm
  M}_\odot$ isolated cluster. With increasing $\delta M_{\rm BH}$
  bubble feedback becomes more infrequent, but also more violent,
  leading to a substantial heating of the central gas.}
\label{sijackiF4}  
\end{figure}

We have performed simulations of isolated galaxy clusters for a range
of different masses, from $10^{13}\,h^{-1} {\rm M}_\odot$ to
$10^{15}\,h^{-1} {\rm M}_\odot$, analyzing the black hole growth and
feedback over a large time span. The initial conditions have been set
up by imposing hydrostatic equilibrium for the gas within a static NFW
dark mater halo, and by introducing a seed black hole sink particle
with $10^{5}\,h^{-1} {\rm M}_\odot$. The simulations have been evolved
for $0.25t_{\rm Hubble}$ with radiative cooling, star formation, and
AGN feedback. In the left-hand panel of Fig.~\ref{sijackiF3} we show
how the black hole mass is growing for three halos of increasing
mass. After the initial rapid growth, bubble feedback regulates the
black hole mass, which remains practically constant for more than a
Gyr of the simulated time. The mass accretion rate onto the black hole
(see right-hand panel of Fig.~\ref{sijackiF3}) drops after the initial
phase to very low values of order $10^{-3}M_{\rm Edd}$. It can been
seen that the black hole accretion rate shows occasional bursts during
short time intervals, reflecting the recurrent nature of the bubble
feedback. However, these jumps in black hole accretion rate do not
contribute significantly to the growth of the black hole itself.
In Fig.~\ref{sijackiF4} we plot entropy and temperature profiles of a
$10^{14}\,h^{-1} {\rm M}_\odot$ cluster at $0.25t_{\rm Hubble}$. We
compare the runs without AGN heating with the simulations with
different thresholds $\delta M_{\rm BH}$ for the bubble triggering. In
the case of highest $\delta M_{\rm BH}$, the bubble frequency is
lowest, but the energy injected into the bubbles is large. This model
corresponds to a sporadic, but rather powerful AGN activity and heats
the ICM very efficiently. On the other hand, low values of $\delta
M_{\rm BH}$ imply more frequent and gentle bubble feedback that
affects the ICM properties mildly, but that can still prevent the
overcooling in the central regions of clusters.

\section{Conclusions}
\label{sijacki:Conclusions}
Unless heavily suppressed by magnetic fields, physical gas viscosity
in hot, massive clusters appears to be an important physical
ingredient, changing significantly the properties of AGN-driven
bubbles and influencing the dynamics of clusters during merging
events. The ever more accurate X-ray data on galaxy clusters will
allow detailed comparisons with simulations, that can constrain the
effective level of viscosity present in these systems. AGN heating is
a very promising candidate to solve the cooling flow riddle in
clusters. Adopting the theoretical model outlined here we discuss in
forthcoming work \cite{sijacki:SijackiSpringel2006c} fully
cosmological simulations of self-regulated AGN feedback, trying to
understand this physical mechanism in more depth.
   
\subparagraph{Acknowledgments.} We are grateful to Simon White, Eugene
  Churazov and Andrea Merloni for many constructive discussions and
  comments.

%
%




\printindex
\end{document}